\documentclass{nature}
\usepackage{graphicx}
\usepackage{endfloat}
\usepackage{amsmath}
\usepackage{amsfonts}

\title{Directing nanoscale optical flows by coupling photon spin to plasmon extrinsic angular momentum}

\author{Y. Lefier, R. Salut, M.A. Suarez, T. Grosjean*}

\begin{document}

\maketitle

\begin{affiliations}
\item FEMTO-ST Institute, Universit\'e  Bourgogne Franche-Comt\'e, UMR CNRS 6174 15B Av. des Montboucons, 25030 Besancon cedex, France
\end{affiliations}

\begin{abstract}

As any physical object, light undergoing a circular trajectory features a constant extrinsic angular momentum. Within strong curvatures, this angular momentum can match the spin momentum of a photon, thus providing the opportunity of a strong spin-orbit interaction. Using this effect, we demonstrate tunable symmetry breaking in the coupling of light in a curved nanoscale plasmonic waveguide. The helicity of the impinging optical wave controls the power distribution between the two counter-propagating guided modes, including unidirectional waveguiding. We found that up to 95 \% of the incoupled light can be directed into one of the two propagation directions of the waveguide. This approach offers appealing new prospects for the development of advanced, deeply subwavelength optical functionalities.

\end{abstract}

\newpage

In addition to its energy, light possesses polarization and spatial degrees of freedom, manifested by its linear momentum as well as spin and orbital angular momenta\cite{allen:pra92}. Remarkably, spin and orbital momenta are not independent quantities; the spin (circular polarization) can tailor the spatial distribution and propagation direction of light \cite{bliokh:natphot15}. This phenomenon, known as spin-orbit interaction (SOI), has recently attracted much interest for applications involving light manipulation \cite{bliokh:prl06,petersen:sci14,rodriguez:prl10,bomzon:ol01,shitrit:nl11,bliokh:prl08,brasselet:prl13,gorodetski:prl13}.

In particular, the ability of SOI to control the directionality of guided waves, such as planar surface plasmons \cite{lin:sci13,xi:joo14,rodriguez:sci13,mueller:nl14,shitrit:sci13,huang:lsa13} and dielectric waveguide modes \cite{petersen:sci14,mitsch:natcom14,feber:natcom15}, holds promise of an unprecedented level of control of the flow of light in advanced integrated systems and networks. So far, SOI has enabled the unidirectional waveguiding of diffraction-limited modes. The control of the propagation direction within nanoscale plasmonic waveguides \cite{gramotnev:natphot10,huang:nl09,kriesch:nl13,lefier:ol15} thus remains an important challenge.

Here, we introduce the concept of an SOI based on the conversion of the spin angular momentum (SAM) of a diffraction-limited incoming wave to the extrinsic orbital angular momentum (EOAM) of a guided mode within a sharply curved waveguide. This concept is demonstrated with the helicity-controlled tunable unidirectional coupling of light directly within a bent gap-plasmon waveguide \cite{pile:apl05a,veronis:ol05,liu:ox05}. Depending on the sign of their SAM, the incoming photons are shown to couple upstream or downstream, with at most 95\% of the incoupled energy which is directed towards one of the two propagation directions. This SOI is used for distributing photons towards resonant nano-antennas, thus opening new routes in advanced optical functionalities based on truly subwavelength integrated plasmonic circuits \cite{huang:nl09,kriesch:nl13}.

\begin{figure}[h!]
\centering
\includegraphics[width=0.6\linewidth]{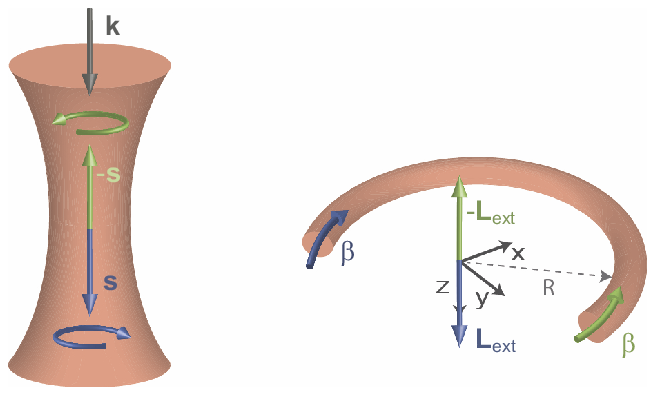}
\caption{\textbf{Conditions for SAM-to-EOAM interaction.} Longitudinal SAM (\textbf{s}) of a freely propagating wave along (0z)-axis together with the EOAM ($\mathbf{L}_{ext}$) of a mode guided along a curved trajectory in the (xy)-plane. Depending on both the polarization handedness of the freely propagating wave and the propagation direction of the guide mode, the SAM and EOAM are parallel or antiparallel. Therefore, if the guided mode is free from intrinsic AM, the SOI between the incident wave and the guided mode is favorable to unidirectional waveguiding. This condition requires a trajectory curvature satisfying $R=\lambda_{mode} /2 \pi$ (Eq. \ref{eq1})} \label{fig:scheme}
\end{figure}

Our approach relies on the transfer of angular momentum (AM) from an optical wave freely propagating along (0z)-axis, towards a guided mode circulating along a curved trajectory within the perpendicular (xy)-plane (Fig. \ref{fig:scheme}). We focused our attention on the waveguide modes of negligible intrinsic AM. In this case, a curved single-mode waveguide sustains two counter propagating modes whose amplitudes have the azimuthal angular dependence $\exp(\pm i\beta R \phi)$. Here, $\beta=2\pi/\lambda_{mode}$ is the wave vector and $\lambda_{mode}$ is the effective wavelength. Analogy with quantum mechanics\cite{marcuse:book} suggests that such modes are the eigenmodes of the angular momentum operator $L_z=-i\hbar\partial/\partial \phi $, and thus carry extrinsic AM of $\pm \hbar \beta R$  per photon along (0z)($\hbar=h/2\pi$ refers to the Planck constant). The incoupled light being the superposition of these two modes, its EOAM equals $(|a_r|^2-|a_l|^2) \hbar \beta R$ per photon, where $|a_r|^2$ and $|a_l|^2$ are the probabilities that a guided photon propagates along the two opposite right and left directions, respectively. An impinging plane wave propagating along (0z) carries a SAM  of $\sigma \hbar$ per photon oriented along (0z), where $\sigma$ defines polarization helicity. In the case of an "`elastic"' coupling between the incident photons and a guided mode within an axisymmetrical (circular) waveguide, the longitudinal AM is conserved and the probabilities $|a_r|^2$ and $|a_l|^2$ reads:

\begin{eqnarray}
 |a_r|^2=\frac{1+\sigma/\beta R}{2}, \label{eq:p1} \\
 |a_l|^2=\frac{1-\sigma/\beta R}{2}. \label{eq:p2}
\end{eqnarray}

For a linearly polarized incoming wave ($\sigma$=0) or a rectilinear waveguide ($R\rightarrow\infty$) light coupling is symmetric as $|a_r|^2=|a_1|^2=1/2$. Unidirectionality (ie, $|a_r|^2=1$ and $|a_l|^2=0$, or vice versa) arises when the SOI satisfies the geometricodynamical condition:

\begin{equation}
 \beta =  \frac{\sigma}{R}. \label{eq1}
\end{equation}

Equations \ref{eq:p1} and \ref{eq:p2} thus unveal a trajectory curvature on the subwavelength scale which leads to tunable unidirectional mode coupling. Like in the optical spin hall effect \cite{bliokh:prl06}, the SAM-to-EOAM conversion is closely related to a geometric phase, whose gradient is here equal to $\nabla \Phi_G=\sigma/R$. Equation \ref{eq1} can thus be regarded as the geometric-to-dynamic phase matching originating the unidirectionality.

Equation \ref{eq1} requires identifying tightly confined guided modes strongly bound to the waveguide, to limit mode leakage in sharply curved trajectories. To fulfill this condition, we focused our attention on the so-called gap-plasmon waveguide design \cite{pile:apl05a,veronis:ol05,liu:ox05}. This geometry supports a nanometer scale guided mode whose field distribution is transversely polarized \cite{veronis:ol05}, making it a good candidate for producing the desired geometric phase gradient $\nabla \Phi_G=\sigma/R$ \cite{bliokh:prl08}. The real conditions also involve an impinging focused beam, rather than a plane wave, and a non axisymmetric lossy waveguide, which may affect the AM conservation introduced above. To further assess surface plasmon directionality in real conditions, we implemented an analytical wave model, based on our SOI-induced geometrical phase $\Phi_G$, to describe the gap plasmon excitation under illumination with a circularly polarized focused beam (see details in the supplementary material S1.1). Our model showed that under real conditions, unidirectionality is still effective but spectrally shifted with respect to the predictions from Eq. \ref{eq1} (supplementary material S1.2).

We performed full-wave finite-difference time domain (FDTD) simulations of the nanoscale guided mode created by a bent gap-plasmon waveguide of 180$^{\circ}$ curvature angle patterned on a gold surface, for various incident polarizations. This structure was designed for operation at $\lambda = 1.55$ $\mu$m, corresponding to $\lambda_{mode}$=981 nm and a propagation length $L=6.67$ $\mu$m. The geometrical parameters of the structure are R = 500 nm and $w=50$nm (gap width), and the thickness of the gold film is 100 nm. The results agree well with our analytical model, in terms of directionality and field distribution within the bent plasmonic waveguide (see supplementary material S1.2). The slightly lower directionality observed with FDTD simulations (94\% with FDTD versus 100 \% with our model) shows that any residual excitation on both sides of the waveguide bend leads to a loss of directionality. Our model considers excitation only in the waveguide curvature.

\begin{figure}[h!]
\centering
\includegraphics[width=0.95\linewidth]{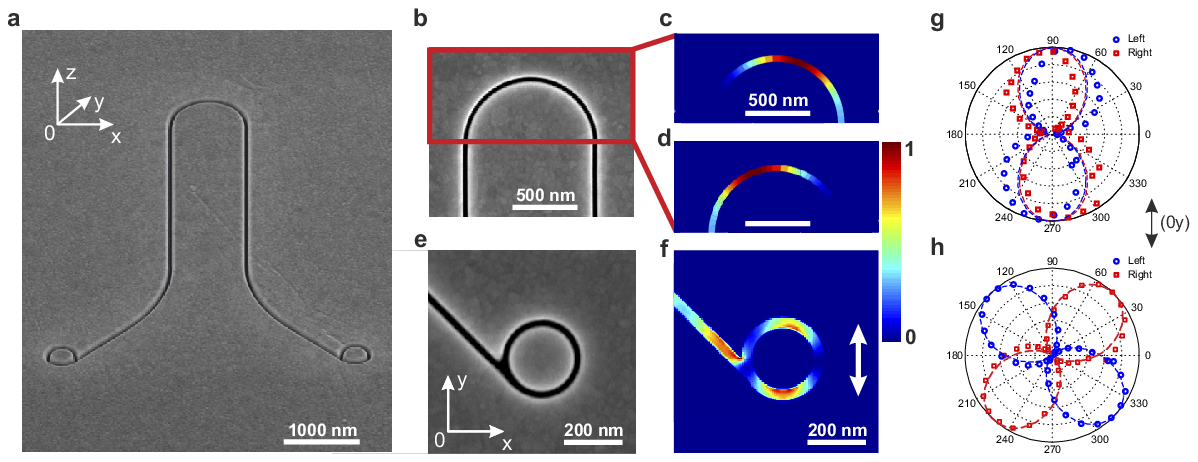}
\caption{\textbf{A nanoscale optical transmission-line enabling selective near-field excitation of two resonant coaxial nano-antennas by spin-orbit interaction.} (a) Scanning electron micrograph of a subwavelength integrated optical circuit consisting of a bent gap plasmon waveguide connected to coaxial nano-antennas. The structure was fabricated in a gold film for operation at $\lambda$ = 1.55 $\mu$m. (b) Zoom on the 180$^{\circ}$ waveguide curvature. (c,d) Simulated electric field intensity within the curved gap-plasmon waveguide  under illumination at $\lambda=$ 1.55 $\mu$m, with our analytical model. The incoming gaussian beam is 1 $\mu$m wide, centered to the top of the bend and features (c) circular right and (d) circular left polarization. Our model considered a mode excitation only in the waveguide bend. (e,f) Zoom on the plasmon-fed coaxial nano-antenna: (e) SEM image of the nano-antenna (f) Simulated electric field intensity within the waveguide and the nano-antenna, with FDTD method. The nano-antenna shows a dipole moment along the white arrow, ie. parallel to the (0y)-axis. (g) Experimental emission intensity on the left (blue circle) and right (red squares) nano-antennas as a function of the angle $\theta$ with respect to (0x)-axis of a linear analyzer placed before the camera. The waveguide bend is excited with a linearly polarized focused beam so that the two nano-antennas radiate at the same time. The results are compared to theoretical predictions (dashed blue and red curves, respectively). (h) Evolution of the experimental emission intensity at the two ends of the same circuit without nano-antennas, in a configuration similar to (g). The infrared camera used for measurements shows a linear response over its entire dynamic range.}
\label{fig:exp1}
\end{figure}

We fabricated a bent gap plasmon waveguide of width $w=50$nm in a 100-nm-thick gold film using focused ion beam milling (FIB)(Fig. \ref{fig:exp1}(a)). On both sides of the curvature, of radius R=500 nm (Fig. \ref{fig:exp1}(b)), the waveguide expands over approximately 4.5  $\mu$m (3 $\lambda$) and the two channels are tilted by $45^{\circ}$ after 2 $\mu$m to extend the distance between the two waveguide ends to about 2 $\lambda$. Two plasmonic coaxial nano-antennas\cite{baida:prb06} were milled by FIB at the two waveguide ends to resonantly concentrate optical fields and fully control free-space emission (Fig. \ref{fig:exp1}(e))\cite{huang:nl09,kriesch:nl13}. The inner radius of the coaxial nano-antennas has been set to 150 nm to spectrally center their resonance to 1550 nm. FDTD simulations showed that the emission of these nano-antennas is governed by a electric dipole moment oriented along the (0y)-axis (Fig. \ref{fig:exp1}(f) and (g)).

The curved portion of the waveguide was back illuminated with polarized 1550-nm laser light in a continuous wave regime, and the overall device was imaged in transmission mode using a far-field microscope coupled to an infrared camera. The polarization of the incoming beam was adjusted with a linear polarizer coupled to a quarter-wave plate. Depending on the angle $\alpha$ between the respective element axis, the transmitted field varied from linear ($\alpha=0^{\circ}$, $90^{\circ}$) to circular ($\alpha=45^{\circ}$, $135^{\circ}$) polarizations, with elliptic polarizations for intermediate angles. First, the optical emission of the two fabricated nano-antennas excited  with linearly polarized light has been analyzed  by positioning a rotating analyzer before the camera.  In accordance with the simulations, the emission intensity of both nano-antennas is maximum when the polarizer is aligned with the (0y)-axis (Fig. \ref{fig:exp1}(g)).  Without nano-antennas, the detected signal is, as expected, maximum when the analyzer is aligned across the waveguide's gap (at 45$^{\circ}$ with respect to (0y)-axis, Fig. \ref{fig:exp1}(h)).

\begin{figure}[h!]
\centering
\includegraphics[width=0.9\linewidth]{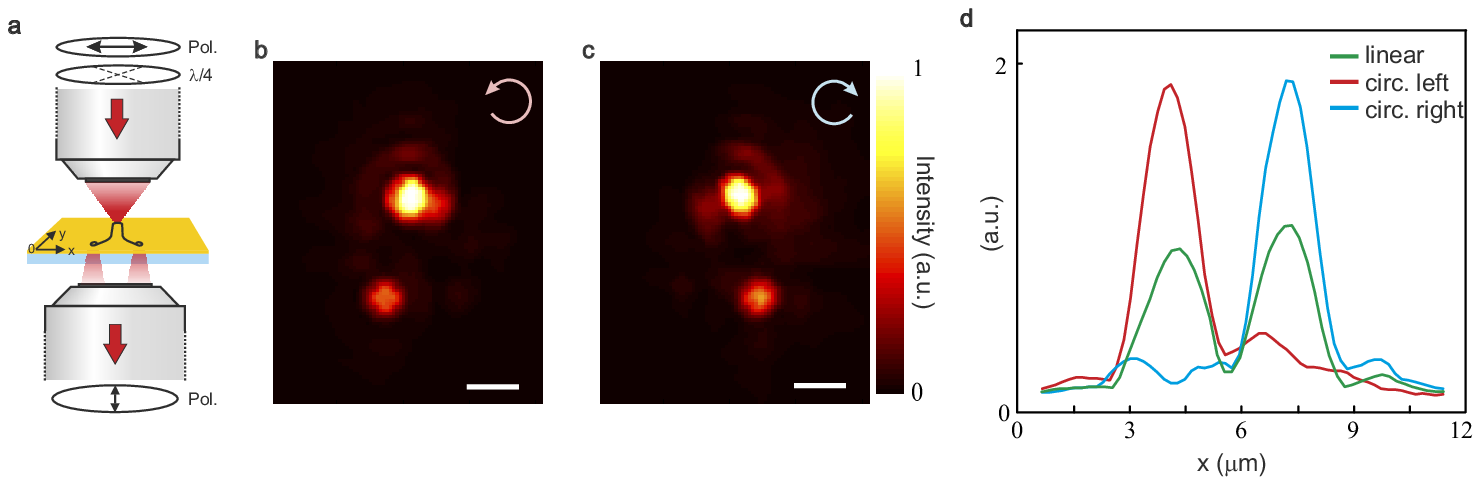}
\caption{\textbf{Spin-orbit interaction directs surface plasmons on the nanometer scale.} (a) Schematic diagram of the optical set-up used for characterizing the surface plasmon directionality within our nanoscale network. Directional surface plasmon excitation is carried out with an illumination of the waveguide bend from the front with a circularly polarized beam focused onto the top of the curvature with a ($\times 60, 0.9$) objective. The overall structure is imaged in the far-field from the back with a ($\times 125,0.7$) objective coupled to a polarizer and a camera. (b,c) Optical emission from the structure observed with (b) left circular and (c) right circular polarizations. The state of incident polarization is shown by red and blue arrows, respectively. (d): intensity profile across the two nano-antennas with left circular (red curve), right circular (blue curve) and linear (green curve) excitation polarization. The linear polarization is parallel to the (0x)-axis.}
\label{fig:exp2}
\end{figure}

The structure was then sandwiched between a circular polarizer (i.e. a linear polarizer and a quarter-wavelength retardation plate) and a linear polarizer (Fig. \ref{fig:exp2}(a)). The observed field emissions (Figs. \ref{fig:exp2}(b) and (c)) agree well with the surface plasmon directionality theoretically predicted with our analytical model (Fig. \ref{fig:exp1}(c) and (d)). Circularly polarized light beams of opposite handedness generate localized optical emission at either the right or the left coaxial nano-antenna whereas linearly polarized light induces hot spots of equal intensity profiles at both nano-antennas (Fig. \ref{fig:exp2}(d)). 
The coupling directionality was estimated from the relative nano-antenna emission intensities deduced from Figs. \ref{fig:exp2}(b) and (c). In accordance with FDTD simulations, the directionality for both circular polarizations is equal to 95 \%.

\begin{figure}[h!]
\centering
\includegraphics[width=0.7\columnwidth]{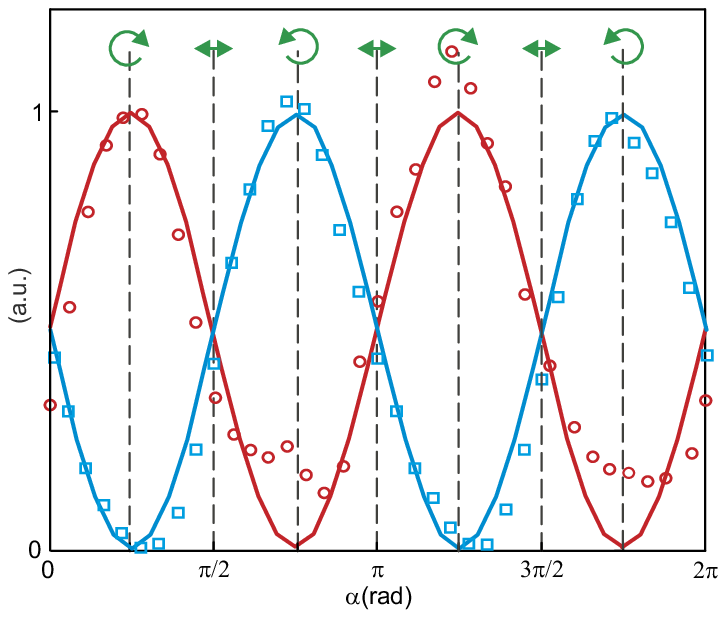}
\caption{\textbf{Changing the incident polarization enables continuous tuning of the directionality of the launched nanoscale guided mode.} Evolution of the emission intensities of the two fabricated nano-antennas as a function of the polarization of the incoming beam. The incident polarization is set by the angle $\alpha$ between a polarizer and a quarter wavelength plate in the illumination system. It varies continuously from linear state ($\alpha=$0, $\pi/2 $) to left and right circular states ($\alpha=\pi/4$ and $3\pi/4$, respectively) with intermediate elliptic states.}
\label{fig:exp3}
\end{figure}

The emission intensities of the nano-antennas are shown in Fig. \ref{fig:exp3} as a function of the polarization of the incident light. The polarization is set by the angle $\alpha$ between the polarizer and the quarter wavelength plate in the illumination system. These results agree well with the polarization-dependent distribution of incoming photons in the spin states "right" and "left", whose respective intensities $|\psi_r|^2$ and $|\psi_l|^2$ read:

\begin{eqnarray}
|\psi_r|^2 & = & \frac{1}{2} (1+ \sin 2 \alpha),\\ \label{eq4}
|\psi_l|^2 & = & \frac{1}{2} (1- \sin 2 \alpha), \label{eq5}
\end{eqnarray}

the helicity of the incoming beam being defined as $\sigma=|\psi_r|^2-|\psi_l|^2= \sin 2 \alpha$. For all angles $\alpha$, the incident photons prepared in the right and left spin states are selectively directed to the nano-antennas on the right and left, respectively, leading to a directional coupling tunable with polarization.

In conclusion, we demonstrated the tunable excitation of resonant nano-antennas by the spin-controlled unidirectional coupling of surface plasmons within a nanometer-scale metallic waveguide.  Our approach, based on the coupling between the SAM of light and the EOAM of guided wave surface plasmons, opens new perspectives in manipulating deeply subwavelength optical fields within compact systems made of interconnected nano-optical elements. More generally, SOI within nanoscale devices has the potential of becoming a key building block of future nano-optical technologies.

\begin{addendum}
\item The authors are indebted to Ulrich Fischer for helpful discussions, and to Cara Leopold for her support in the paper writing. This work is supported by the Labex ACTION program (contract ANR-11-LABX-01-01), and by the French RENATECH network and its FEMTO-ST technological facility.
\item[Competing Interests] The authors declare that they have no competing financial interests
\item[Author contributions] Y.L. and T.G. conceived the idea and designed the study. Y.L and T.G. contributed to the theoretical and numerical studies. R.S. fabricated the structures. Y.L, M.A.S and T.G. performed the experiments and data analysis. T.G. wrote the paper and supervised the project. All co-authors approved the final version of the manuscript.
\item[Correspondence] Correspondence and requests for materials should be addressed to T.G.\\ (email: thierry.grosjean@univ-fcomte.fr).
\end{addendum}


\end{document}